\documentclass[preprintnumbers,11pt,a4paper,oneside]{article}
\pdfoutput=1

\usepackage{contour}
\usepackage{xcolor}

\usepackage{mathrsfs}

\contourlength{0.05em} 

\usepackage{jheppub}
\usepackage{color}
\usepackage{graphicx}
\usepackage{wrapfig,enumerate,slashed}

\usepackage{footmisc}
\usepackage{amsmath}
\usepackage{wasysym} 
\usepackage{graphicx}
\usepackage{subcaption}
\usepackage{color}
\usepackage{comment}
\usepackage{appendix}
\usepackage{slashed}
\usepackage{siunitx}

\newcommand{\ket}[1]{\left| #1 \right>} 
\newcommand{\bra}[1]{\left< #1 \right|} 
\newcommand{\braket}[2]{\left< #1 \vphantom{#2} \right|
	\left. #2 \vphantom{#1} \right>} 
\newcommand{\matrixel}[3]{\left< #1 \vphantom{#2#3} \right|
	#2 \left| #3 \vphantom{#1#2} \right>} 

\usepackage{tikz}
\usepackage{orcidlink}

\makeatletter
\gdef\@fpheader{}
\makeatother
\begin{document}

\title{Real-Time Scattering on Quantum Computers via Hamiltonian Truncation}

\author[a]{James~Ingoldby,\orcidlink{0000-0002-4690-3163}}
\author[a]{Michael Spannowsky,\,\orcidlink{0000-0002-8362-0576}}
\author[a]{Timur Sypchenko,\,\orcidlink{0009-0008-3894-7458}}
\author[a]{Simon Williams\,\orcidlink{0000-0001-8540-0780}}
\author[b]{Matthew Wingate\,\orcidlink{0000-0001-6568-988X}}

\emailAdd{james.a.ingoldby@durham.ac.uk}
\emailAdd{michael.spannowsky@durham.ac.uk}
\emailAdd{timur.sypchenko@durham.ac.uk}
\emailAdd{simon.j.williams@durham.ac.uk}
\emailAdd{m.wingate@damtp.cam.ac.uk}

\affiliation[a]{\vspace{0.1cm} Institute for Particle Physics Phenomenology, Durham University, Durham DH1 3LE, UK}
\affiliation[b]{\vspace{0.1cm} DAMTP, University of Cambridge, Cambridge, CB3 0WA, UK}

\preprintA{IPPP/25/24}

\abstract{We present a quantum computational framework using Hamiltonian Truncation (HT) for simulating real-time scattering processes in $(1+1)$-dimensional scalar $\phi^4$ theory. Unlike traditional lattice discretisation methods, HT approximates the quantum field theory Hilbert space by truncating the energy eigenbasis of a solvable reference Hamiltonian, significantly reducing the number of required qubits. Our approach involves preparing initial states as wavepackets through adiabatic evolution from the free-field theory to the interacting regime. We experimentally demonstrate this state preparation procedure on an IonQ trapped-ion quantum device and validate it through quantum simulations, capturing key phenomena such as wavepacket dynamics, interference effects, and particle production post-collision. Detailed resource comparisons highlight the advantages of HT over lattice approaches in terms of qubit efficiency, although we observe challenges associated with circuit depth scaling. Our findings suggest that Hamiltonian Truncation offers a promising strategy for quantum simulations of quantum field theories, particularly as quantum hardware and algorithms continue to improve.}
\maketitle


\section{Introduction}

The simulation of scattering processes in quantum field theories (QFTs) presents a foundational and computationally demanding challenge, central to advancing our understanding of fundamental interactions and particle phenomenology. Classical approaches are typically rooted in Monte Carlo algorithms, which suffer from the so-called \emph{sign-problem}~\cite{PhysRevLett.46.77, VONDERLINDEN199253, PhysRevE.49.3855}, severely limiting their applicability to regimes involving real-time dynamics or finite chemical potential. Quantum computers have the ability to efficiently simulate the real-time evolution of highly-entangled quantum systems without such limitations and therefore provide a natural framework in which to study QFTs in these regimes. Recent advances in quantum algorithms have demonstrated the promise of such simulations~\cite{Klco:2018zqz, Banuls:2019bmf, Liu:2020eoa, PRXQuantum.4.027001, Kane:2022ejm, crane2024}, opening new avenues for studying problems in high-energy physics that are intractable with classical methods. These include real-time scattering dynamics~\cite{Jordan:2011ci, Jordan:2012xnu, Jordan:2014tma, Martinez:2016yna, Jordan_2018, Araz:2024kkg, zemlevskiy2024, Chai2025fermionicwavepacket, Abel:2025zxb}, thermalisation~\cite{PhysRevD.106.054508, doi:10.1126/science.abl6277, PhysRevA.108.022612, Fromm:2023npm, Araz:2023ngh}, and non-equilibrium dynamics~\cite{PhysRevLett.109.175302, Abel:2020qzm, Abel:2020ebj, Angelides:2023noe, Ingoldby:2024fcy}.

In this paper, we focus on the simulation of real-time scattering processes in the (1+1)-dimensional $\phi^4$ scalar field theory. Following the seminal work of Jordan, Lee, and Preskill~\cite{Jordan:2011ci, Jordan:2012xnu}, most approaches to simulating QFTs on quantum devices employ a lattice discretisation procedure to approximate the theory's Hilbert space. In contrast, we propose a Hamiltonian Truncation (HT) framework~\cite{fitzpatrick2022,Konik-review}, which generalises the Rayleigh–Ritz variational method by truncating the energy eigenbasis of a solvable reference Hamiltonian.

We demonstrate that the HT approach offers several advantages over conventional lattice-based methods. In particular, constraints such as momentum conservation can be imposed from the start in constructing the truncated basis. By enforcing momentum conservation at this stage, states with incompatible momentum are excluded, reducing both the size of the Hilbert space and the number of qubits required to simulate scattering processes. Moreover, the ground state of the free QFT is constructed to correspond exactly to the zeroth state in the computational basis of the qubit-based device, eliminating the need for complex ground state preparation routines~\cite{Ingoldby:2024fcy}. The HT framework is therefore particularly well suited to near-term devices which are constrained by small numbers of qubits and short coherence times. 

We introduce a quantum algorithm tailored to compute the real-time dynamics of scattering states within the HT framework. We present a method for the preparation of initial wavepacket states, achieved via adiabatic evolution from the free theory to the interacting theory, an approach known as \textit{adiabatic state preparation}~\cite{farhi2000quantumcomputationadiabaticevolution, RevModPhys.90.015002, 10.1063/1.2798382, 10.1063/1.4748968, Jordan:2011ci}. To demonstrate the practical applicability of this method, we validate our state preparation procedure on the IonQ Aria 1 quantum computer, a 25-qubit trapped-ion device based on $^{171}\textrm{Yb}^+$ ions~\cite{Wright_2019}.

To further assess the suitability of the HT framework for NISQ-era devices, we carry out an assessment of the scalability and resource demands of our framework, performing a detailed comparison with conventional lattice-based approaches. Our results show that HT substantially reduces the number of required qubits; however, this comes at the cost of increased circuit depth, which scales exponentially with the truncation energy. We conclude by discussing prospective algorithmic improvements aimed at mitigating this depth scaling and enhancing the overall viability of the method for near-term quantum hardware.

The structure of this paper is as follows: Section~\ref{sec:HTSection} outlines the Hamiltonian truncation method, detailing our choice of basis and the computation of matrix elements. In Section~\ref{sec:scatteringSection}, we elaborate on the critical task of state preparation, with a focus on the adiabatic construction of interacting wavepackets. We also discuss the extraction of physically relevant observables, such as particle production probabilities, and present results from real-time scattering simulations using a quantum emulator. Section~\ref{sec:QCRunSection} presents demonstrations of adiabatic state preparation on real quantum hardware. Section~\ref{sec:resourcesSec} addresses the computational resources required for our simulations, providing a detailed comparison with conventional lattice-based approaches. Finally, Section~\ref{sec:conclusion} summarises our findings and outlines future directions for quantum simulations of quantum field theories.


\section{Hamiltonian Truncation}\label{sec:HTSection}


To approximate our QFT with a finite quantum system suitable for analysis on a quantum computer, we employ Hamiltonian Truncation (HT). In this approach, which generalises the Rayleigh-Ritz method from quantum mechanics, the original Hamiltonian $H$ is first decomposed into a solvable piece $H_0$ plus an interaction $V$. Then the Hilbert space basis in our approximation is constructed by enumerating all the lowest energy eigenstates of the solvable Hamiltonian satisfying $H_0\ket{i} = E_i\ket{i}$ with $E_i\le E_T$ for a given cutoff energy $E_T$. The approximate Hamiltonian for the full system then takes the form of an explicit matrix of finite dimensionality with elements given by 
\begin{align}
	\matrixel{i}{H}{j} = E_i\delta_{ij} + \matrixel{i}{V}{j}\,.
	\label{eq:htham}
\end{align}
See Refs.~\cite{fitzpatrick2022,Konik-review} for reviews of the Hamiltonian Truncation approach.

As the energy cutoff $E_T$ and size of the truncated basis increases, the truncated Hamiltonian is expected to provide a progressively better approximation of the low-energy states of the original system. This convergence can be systematically improved by constructing an effective Hamiltonian for the QFT, which mitigates truncation effects and accelerates the approach to the large-$E_T$ limit~\cite{Cohen:2021erm}. However, when the original system is a QFT, UV regularization and renormalization must be performed before taking $E_T$ large to ensure that all extracted quantities remain free of UV divergences and converge to well-defined values~\cite{EliasMiro:2022pua,Delouche:2023wsl}.


\subsection{$\phi^4$ on the circle}\label{sec:phi4HT}

We consider the quantum field theory of a single real scalar field $\phi$ in $1+1$ dimensions with a quartic self-interaction, which has been investigated using a variety of Hamiltonian Truncation (HT) approaches in References~\cite{Rychkov:2014eea, Coser:2014lla,Rychkov:2015vap, Bajnok:2015bgw,Elias-Miro:2015bqk,Elias-Miro:2017tup,Elias-Miro:2017xxf, Cohen:2021erm,Szasz-Schagrin:2022wkk, Lajer:2023unt,Harindranath:1987db,Burkardt:2016ffk,Anand:2017yij,Chen:2021bmm,Chen:2021pgx}. The Lagrangian density takes the form
\begin{align}
	\mathcal{L} = \frac{1}{2} \left(\partial_\mu \phi\right)^2 - \frac{1}{2} M^2 \phi^2 - g \phi^4~.
\end{align}
To analyse this theory within the framework of HT, we place the theory on a circle of length $L$, enforce periodic boundary conditions on the field and apply canonical quantisation on equal-time slices. The free Hamiltonian is defined as
\begin{align}
	H_0 = \frac{1}{2} \int_0^L dx\, :\Pi^2 + \left(\partial_x \phi\right)^2 + M^2 \phi^2:~,
\end{align}
where normal ordering, denoted by $:\,:$, is performed with respect to the free vacuum so that the condition $\langle 0 \vert H_0 \vert 0\rangle = 0$ is satisfied. By expanding the field in terms of creation and annihilation operators,
\begin{align}
	\phi(x) = \sum_{n=-\infty}^{\infty} \frac{1}{\sqrt{2L\omega_n}} \left(a_n e^{i k_n x} + a_n^\dagger e^{-i k_n x}\right)~,
\end{align}
where $k_n = \frac{2\pi n}{L}$ and $\omega_n = \sqrt{k_n^2 + M^2}$, the free Hamiltonian may be rewritten in the diagonal form
\begin{align}
	H_0 = \sum_{n=-\infty}^{\infty} \omega_n a_n^\dagger a_n~,
	\label{eq:H0}
\end{align}
where the creation and annihilation operators satisfy the equal-time bosonic commutation relations, 
\begin{align}
	[a_n, a_m] = [a_n^\dagger, a_m^\dagger] = 0, \quad [a_n, a_m^\dagger] = \delta_{n,m}~.
	\label{eq:comms}
\end{align}

The truncated basis is constructed from the eigenstates of $H_0$, which are Fock states of the form
\begin{equation}\label{eq:basisstates}
	\ket{\{\bf r\}} = \prod_{n=-\infty}^{\infty} \frac{1}{\sqrt{r_n!}} \left(a_n^\dagger\right)^{r_n} \ket{0}~,
\end{equation}
where $\ket{0}$ is the vacuum state satisfying $a_n \ket{0} = 0$ for all modes $n$. In constructing our basis, we arrange the eigenstates in order of increasing $H_0$ eigenvalue and truncate so that only a finite number of states with the smallest energies are included. 

Furthermore, we use the symmetries and selection rules of the $\phi^4$ theory to further restrict the states included in the basis. Both total momentum and parity are conserved quantities in scattering processes on the circle. As a result, we construct our basis to include only states that have zero total momentum and are even under parity transformations. Since the interaction term $V$ only couples states with the same quantum numbers, restricting our basis in this way ensures that we capture the relevant subspace of the Hilbert space for studying low-energy scattering processes.

To identify the zero momentum states, we construct the total momentum operator
\begin{equation}
	P = \sum_{n=-\infty}^\infty k_n a^\dagger_n a_n~,
\end{equation}
and select only basis states that are annihilated by $P$. We also need to identify the parity even subsector of the truncated Hilbert space. In general, parity and total momentum are noncommuting operators, so finding simultaneous eigenstates of both is impossible. However, simultaneous eigenstates with \emph{zero} momentum and even parity can be constructed from the states in Equation~(\ref{eq:basisstates}) with the following recipe \cite{Rychkov:2014eea}
\begin{equation}
	\ket{\{\bf \hat{r}\}} = \beta \left(\ket{\{\bf r\}} + \mathbb{P}\ket{\{\bf r\}}\right)~,
	\label{eq:paritystates}
\end{equation}
where the occupation numbers $r_n$ are sent to $r_{-n}$ under the action of the parity operator $ \mathbb{P}$, and $\beta=1/2$ if $\mathbb{P}\ket{\{\bf r\}} = \ket{\{\bf r\}}$ and is $1/\sqrt{2}$ otherwise.

The interacting part of the Hamiltonian is given by
\begin{align}
	V = g \int_0^L dx :\phi^4(x):~.
\end{align}
Using the mode expansion of $\phi(x)$, this interaction term can be rewritten as
\begin{multline}
	V = \frac{gL}{4} \sum_{\substack{n_1 + n_2 + \\
			n_3 + n_4 = 0}} \Bigg[\prod_{i=1}^4(L \omega_{n_i})^{-1/2} \Big( a_{n_1} a_{n_2} a_{n_3} a_{n_4} + 4 a^\dagger_{-n_1} a_{n_2} a_{n_3} a_{n_4} \\[0.5em]
	+ 6 a^\dagger_{-n_1} a^\dagger_{-n_2} a_{n_3} a_{n_4} + 4 a^\dagger_{-n_1} a^\dagger_{-n_2} a^\dagger_{-n_3} a_{n_4} + a^\dagger_{-n_1} a^\dagger_{-n_2} a^\dagger_{-n_3} a^\dagger_{-n_4} \Big)\Bigg]~.
	\label{eq:V}
\end{multline}
The matrix elements of $V$ between basis states of the form in Equation~(\ref{eq:paritystates}) can then be computed using Equation~(\ref{eq:V}) by applying the commutation rules for the bosonic operators shown in Equation~(\ref{eq:comms}). With this foundation in place, we now turn to the construction of wavepacket scattering states within this truncated framework.


\section{Real-time scattering of (1+1)-dimensional $\phi^4$ theory}\label{sec:scatteringSection}

The real-time simulation of scattering processes in quantum field theory (QFT) is a foundational and computationally demanding challenge which is critical to understanding the interactions of fundamental particles. In this section, we use the Hamiltonian Truncation (HT) framework from Section~\ref{sec:HTSection} to construct scattering states on a qubit-based quantum computer using adiabatic state preparation~\cite{farhi2000quantumcomputationadiabaticevolution, RevModPhys.90.015002, 10.1063/1.2798382, 10.1063/1.4748968, Jordan:2011ci} and simulate the real-time dynamics of the quantum fields in $(1+1)$-dimensional $\phi^4$ scalar field theory. We consider the preparation of scattering states, and observe that in the HT formalism the ground state of the QFT corresponds exactly to the ground state of the qubit-based system, therefore removing the need for complicated ground-state preparation routines. This consequently reduces the circuit depths required to simulate scattering processes on the quantum device. Furthermore, it will be shown that the HT formalism allows for efficient encoding of basis states onto the qubit-based quantum device, reducing the total amount of qubits needed to simulate the non-perturbative processes discussed here in comparison to traditional lattice approaches. These advantages illustrate the suitability of the HT formalism to noisy intermediate-scale quantum (NISQ) devices, and are discussed in detail in Section~\ref{sec:resourcesSec}. 

To facilitate the real-time evolution, we must construct a Hamiltonian and Schr\"odinger evolve a quantum state under the unitary evolution operation, such that the state at time $t$ is defined as
\begin{equation}\label{eqn:timeEvolution}
\vert\psi(t, x)\rangle = U(t) \vert \psi(0, x) \rangle\equiv e^{-iHt} \vert \psi(0, x) \rangle~,
\end{equation}
where $\vert \psi (0, x)\rangle$ is the initial state of the system. The unitary time-evolution of a quantum state can be naturally simulated on a quantum computer, however directly computing the evolution for large times is typically infeasible due to the evolution operator becoming exponentially large, and in general dense. Instead, the time-evolution operator can be approximated by employing the Trotter-Suzuki decomposition~\cite{MR0103420, 10.1063/1.526596, PhysRevX.11.011020}, such that 
\begin{equation}\label{eq:utrott}
\mathcal{U} (t) = \left[ \prod_i e^ {- i H_i \delta t} \right]^{t/\delta t}~, 
\end{equation}
where the Hamiltonian has been decomposed into a sum of non-commuting parts, $H = \sum_i H_i$. At first order, this operator approximates the time-evolution from Equation~\eqref{eqn:timeEvolution} up to an error of $\mathcal{O} (\delta t ^2)$, and thus is a good approximation if $\delta t$ is small. From now on, we will refer to this method as \textit{Trotterisation}, and will use the method to simulate the non-perturbative dynamics of a scattering process in $(1+1)$-dimensional $\phi^4$ theory. 


\subsection{Constructing scattering states via Hamiltonian Truncation}\label{sec:statePrep}

We endeavour to simulate a scattering process by computing the real-time evolution of scattering states using the Hamiltonian of an interacting QFT prepared via HT. To do so, we must set up the scattering process on a quantum device by constructing two separated wavepackets in the interacting theory. It has been shown that the preparation of arbitrary quantum states on qubit devices requires exponential circuit depths without ancillary qubits~\cite{Xiaoming10044235}, or polynomial circuit depths at the expense of an exponential number of ancillary qubits~\cite{PhysRevA.83.032302, PhysRevResearch.3.043200, PhysRevLett.129.230504}. A further complication lies in the fact that, for the $\phi^4$ model outlined in Section~\ref{sec:phi4HT}, we are restricted to the zero-momentum, parity-even subsector of the truncated Hilbert space. Here we will outline the construction of wavepackets in the interacting theory by first preparing free-field configurations and adiabatically evolving to the interacting theory.  

As shown in Section~\ref{sec:phi4HT}, the HT framework avoids the need for complex ground state preparation on the quantum device. The model is constructed such that the ground state of the quantum field theory corresponds exactly to the ground state of the quantum computer, namely, the zeroth state in the computational basis, $\vert 0 \rangle$. As a result, one can proceed directly to the preparation of scattering states on the quantum device, without requiring costly ground-state preparation routines.

To begin, let us consider a quantum mechanical wavepacket centred around $x_0$ in position space with a spatial width of $\delta$ and a net momentum $p_0$, such that
\begin{align}
	\psi(x) = A\,e^{-\frac{(x-x_0)^2}{2\delta^2}}\,e^{ip_0x}~.
\end{align}
We impose periodic boundary conditions such that our system is defined on a circle with length $L$. The wavepacket can then be expressed as a superposition of momentum eigenstates,
\begin{align}
\psi_m = \int_{0}^{L}dx \braket{k_m}{x}\braket{x}{\psi}  = \frac{1}{\sqrt{L}} \int_{0}^{L}dx e^{-ik_mx}\psi(x)~,
\end{align}
where the momentum mode $k_m$ is quantised such that $k_m=2\pi m/L$ for $m\in\mathbb{Z}$. To construct well-defined scattering states, it is crucial that the tails of the wavepackets are not clipped by the compactification to the circle. Therefore, by choosing $x_0$ sufficiently far from the boundary and enforcing the limit $\delta \ll L$ the wavepacket takes the form,
\begin{align}
	\psi_m \approx B\,\exp\left({i(p_0-k_m)x_0}\right) \, \exp\left({-\frac{(p_0-k_m)^2\delta^2}{2}}\right)~.
\end{align}

In the free theory, single-particle states can be constructed from quantum mechanical wavefunctions by expressing the wavefunction in the Fock basis
\begin{align}
	\ket{\psi}=\sum_m\,\psi_m a^\dagger_m\ket{0}~,
\end{align}
A wavepacket localised at $x_0 = 0$ with mean momentum $p_0$ can then be written as,
\begin{align}
	\ket{p_0,\,\delta,\,0}=\mathcal{N}\sum_{m=-\infty}^{\infty} \exp\left(-\frac{(p_0-k_m)^2\delta^2}{2}\right)a^\dagger_m\ket{0}~,
	\label{eq:wpckt-origin}
\end{align}
where $\mathcal{N}$ is a normalisation constant. As before, we wish to define a wavepacket centred away from the boundary. This is achieved by applying the translation operator $U(x_0) = \exp\{-iP\cdot x_0\}$ to Equation~\eqref{eq:wpckt-origin}. A wavepacket maximally displaced from the boundary at $x=L/2$ is then given by 
\begin{align}
	\ket{p_0,\,\delta,\,L/2}=\mathcal{N}\sum_{m=-\infty}^{\infty} \exp\left(-i\pi m\right)\,\exp\left(-\frac{(p_0-k_m)^2\delta^2}{2}\right)a^\dagger_m\ket{0}~.
\end{align}
Thus, the effect of the translation is simply to introduce a complex phase.

Ultimately, we want to prepare scattering states in our theory, this is, states composed of two wavepackets, one left-moving and one right-moving. To construct the scattering state of interest, we initialise two wavepackets with equal and opposite momentum expectation values, centred at positions $x_0^{(1)} = 0$ and $x_0^{(2)} = L/2$, respectively. To ensure that neither wavepacket is clipped by the compactification to the circle, we adopt the same construction method as before. First, we build a wavepacket centred at $x=L/2$ and then translate to $x=0$. The resulting two-particle state takes the form
\begin{align}
	\ket{\Psi} = \mathcal{N}\sum_{\substack{m_1=-\infty\\m_2=-\infty}}^{\infty}(-1)^{m_1} \exp\left(-\frac{\delta^2}{2} \left[(p_0-k_{m_1})^2 + (p_0+k_{m_2})^2\right]\right) a^\dagger_{m_1}a^\dagger_{m_2}\ket{0}~.
	\label{eq:2packets}
\end{align}
This state is not an eigenstate of total momentum. However, since the system is defined on a circle, total momentum is conserved, and the Hamiltonian decomposes into block-diagonal sectors labelled by total momentum. Each block induces transitions only between states with the same total momentum. Therefore, one can simplify Equation~\eqref{eq:2packets} by projecting onto the relevant momentum subsectors.

We focus on states in the zero-total-momentum subsector, corresponding to two wavepackets with equal and opposite momentum. The initial scattering state in the free-field theory is then given by
\begin{align}
	\ket{\Psi_0} = \mathcal{N}_0\sum_{m=-\infty}^\infty(-1)^m \exp\left(-(p_0-k_m)^2\delta^2\right)a^\dagger_ma^\dagger_{-m}\ket{0}~.
	\label{eq:initstate}
\end{align}
It is therefore convenient to simplify the sum in Equation~\eqref{eq:initstate} by combining terms that contribute to the same physical state, such that
\begin{multline}
	\ket{\Psi_0} = \mathcal{N}^\prime\Bigg(\sum_{m=1}^\infty(-1)^m\left[e^{-(p_0-k_m)^2\delta^2}+e^{-(p_0+k_m)^2\delta^2}\right]\ket{m,-m}
	+\sqrt{2} \, e^{-p_0^2\,\delta^2}\ket{0,0}\Bigg)~,
	\label{eq:initialstate}
\end{multline}
where the states $\ket{m,-m}$ are taken to be normalised to unity for all $m$, including $m=0$. Note that special care is required for the zeroth mode, since $(a^\dagger_0)^2\ket{0}$ is not normalised to one, hence the factor of $\sqrt{2}$ ensures correct normalisation. A further observation one can make is that each term in Equation~\eqref{eq:initialstate} corresponds to a two-particle state, and therefore only a subset of all states in the truncated basis are included in the sum. This allows for efficient encoding of the state on a quantum device, which will be explained in more detail in Section~\ref{sec:QCRunSection}.

\begin{figure}
\centering
\includegraphics[width=0.8\textwidth]{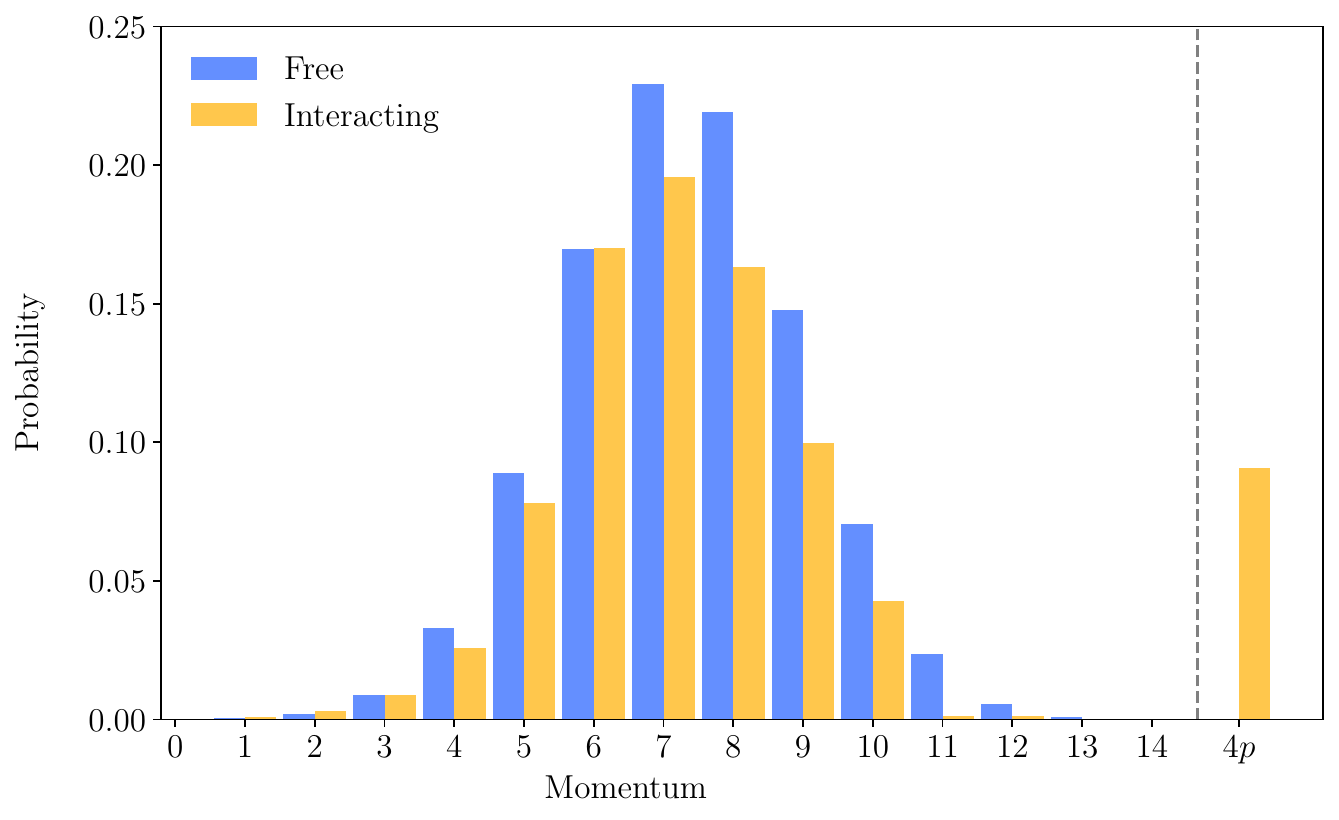}
\caption{Occupation number probabilities of two-particle states for the preparation of free and interacting field wavepackets. The probability of a four-particle state is shown as $4p$. The interacting wavepacket (with coupling $g=1.0$) is prepared using adiabatic state preparation with a ramp time of $t = 1.0$ and time step $ \textrm{d} t = 0.01$. In both cases, we have used a truncation of $n_q = 10$ for an initial momentum of $p_0 = 2.5$ with a momentum spread of $\delta = 0.75$.}\label{fig:wp_sp}
\end{figure}

A final consideration in constructing scattering states in the free theory is to consider the spatial spread, $\delta$, of the wavepackets. We want a value of $\delta$ such that the wavepackets are narrow enough to remain well separated on the circle, yet broad enough that their edges disperse more slowly than the central peak. Therefore, 
\begin{align}
	\frac{1}{p_0} \ll \delta \ll L~.
\end{align}
With this, we are fully equipped to construct scattering states in the free theory. From now on, and for clarity, we will adopt a system of units such that $M=1$. Figure~\ref{fig:wp_sp} shows the occupation number probabilities for a wavepacket constructed in the free field theory using a truncation $n_q = 10$, with initial momentum $p_0 = 2.0$ and a momentum spread $\delta = 0.75$. However, our ultimate goal is to study scattering in the interacting theory. Preparing such states and evolving them in time is classically intractable for all but the smallest and simplest systems. In the following, we outline a procedure to construct wavepackets in the interacting picture on a quantum device, namely adiabatic state preparation. 
 

\subsubsection{Adiabatic state preparation}\label{sec:adiabaticStatePrep}

In the interacting theory, the ground state of the QFT no longer corresponds to a simple product state, and the scattering states of interest cannot be constructed directly from the vacuum. Instead, we require a means for preparing interacting wavepacket states that interpolate smoothly from their free-theory counterparts. A commonly used approach is \textit{adiabatic state preparation}~\cite{farhi2000quantumcomputationadiabaticevolution, RevModPhys.90.015002, 10.1063/1.2798382, 10.1063/1.4748968, Jordan:2011ci}. This method is grounded in the adiabatic theorem, which states that a system initially prepared in an eigenstate of a Hamiltonian, $H_0$, will remain in the corresponding instantaneous eigenstate during a continuous deformation, provided the evolution is sufficiently slow. For the system studied here, this corresponds to constructing wavepackets in the free-theory, which are approximate eigenstates, and then adiabatically ``ramping up" the coupling. This therefore incrementally increases the effect of the interaction term, $V$, from Equation~\eqref{eq:V}, such that the system evolves towards the full interacting Hamiltonian, $H$, over a time $\tau$. If $\tau$ is chosen to be sufficiently long enough to ensure adiabatic evolution, but short enough such that the ramping is completed before the scattering event occurs, the resulting state corresponds to a Gaussian wavepacket in the interacting theory~\cite{Jordan:2011ci}.

Adiabatically preparing the interacting wavepackets requires evolving forward in time by a duration $\tau$, which displaces the wavepackets from their original positions where the free-field states were constructed. To avoid the packets being displaced from their initial positions, we perform the adiabatic ramping up using forward time evolution, and then apply a backward time evolution under the full interacting Hamiltonian to translate the states back to their original spatial locations. The overall state preparation operation therefore has the form,
\begin{equation}
\mathcal{U}_\textrm{SP} =  e^{i \left( H_0 + V\right)  \tau}  \left[\prod_{a=0}^{N} e^{- i \left(H_0 + (1 - \frac{a}{N}) V\right) \delta \tau}\right] 
\end{equation}
where the adiabatic evolution is discretised into $N$ steps, with $\delta \tau = \tau / N$ as adiabatic time step. 

Figure~\ref{fig:wp_sp} shows the occupation number probabilities for a wavepacket constructed in the interacting theory with coupling strength $g = 1.0$ using a truncation of $n_q=10$, an initial momentum $p_0 = 2.5$, and a momentum spread of $ \delta = 0.75$. One can see that, as the interaction is turned on, the wavepacket acquires non-zero occupation number probabilities in the four-particle states when expressed in the free-field basis. To return to the free theory, and thus extract interpretable observables, the system must be ``ramped down", which involves applying the inverse operation of the ramping up process.


\subsection{Examining real-time scattering dynamics}

Having constructed scattering configurations within the HT framework, we now turn our attention to studying the real-time dynamics of these wavepackets to simulate a scattering process. As a testbed, we consider scattering in $(1+1)$-dimensional $\phi^4$ scalar field theory. The scattering states are prepared on a quantum device using the adiabatic state preparation procedure outlined in Section~\ref{sec:statePrep}, followed by Trotterised real-time evolution under the full interacting Hamiltonian. Information about the scattering process is extracted by measuring the expectation values of suitable observables on the quantum device.

\subsubsection{Scattering observables in Hamiltonian Truncation}

To successfully extract scattering data from the simulation, we must define appropriate observables.  To begin, we would like to determine and track the positions of the wavepackets as they propagate through space. For states restricted to the zero-momentum subsector, the uncertainty principle guarantees that the absolute position of a single particle is completely uncertain. However, information about the relative separation between the two wavepackets remains accessible, provided that the system is not prepared in an eigenstate of \emph{relative} momentum. Therefore, to examine this separation during the scattering process, we define the two-particle density operator, 
\begin{align}
	\rho(x_1,x_2)  = \ket{x_1,x_2}\bra{x_1,x_2}~,
\end{align}
which projects onto the component of state corresponding to two particles localised at positions $x_1$ and $x_2$.

It is possible to yield a more convenient form of the two-particle density operator by expressing $\rho(x_1,x_2)$ in the Fock basis and projecting onto the zero-momentum subsector, such that 
\begin{align}
	\rho(y) = \sum_{n,\,m=0}^\infty\ket{n,-n}\phi^*_n(y)\phi_m(y)\bra{m,-m}\,,
\end{align}
where $\phi(y)$ is the wavefunction of two indistinguishable free particles on the circle with length $L$. After the projection, it can be seen explicitly that these wavefunctions, and by extension the density operator, depend only on the separation $y=|x_1-x_2|$. Thus,
\begin{align}
	\phi_n(y) = \begin{cases} 
		\sqrt{\frac{2}{L}} & \text{if } n=0, \\
		\frac{2}{\sqrt{L}}\cos(k_n y) & \text{otherwise }.
	\end{cases}
\end{align}
where the pre-factors ensure that the expectation value of $\rho (y)$ can be interpreted as a properly normalised probability density, such that 
\begin{align}
	1 = \int_0^{L/2}dy \matrixel{\psi_2}{\rho(y)}{\psi_2}\,,
\end{align}
for any two-particle state, $\ket{\psi_2}$.

\begin{figure}
	\begin{center}
		\includegraphics[width=0.95\columnwidth]{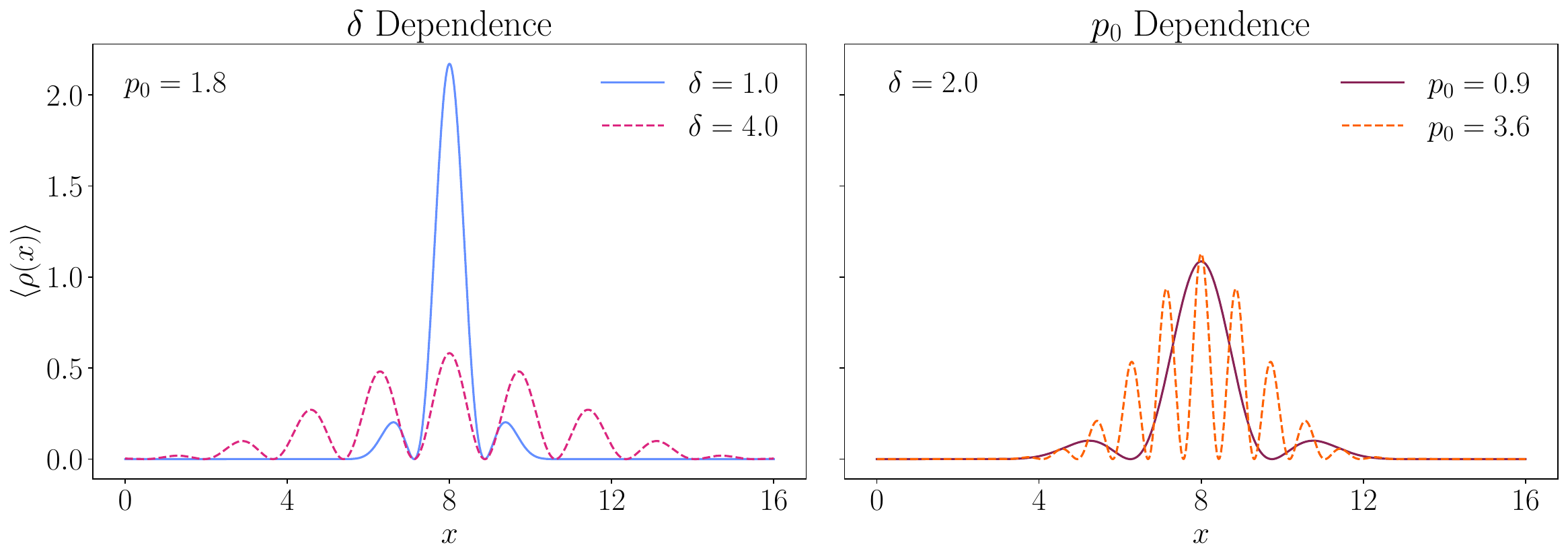}
	\end{center}
	\caption{Expectation value $\langle\rho(x)\rangle$ for wavepacket states on a ring of length $L=16$ with varying initial momenta, $p_0$, and momentum-space spreads, $\delta$. All quantities expressed in units of $M$. The expectation value $\langle\rho(x)\rangle$ can be interpreted as a probability density for the two particles to be separated by a distance $x$.}
	\label{fig:initialwpckt}
\end{figure}

The expectation value of $\rho(x)$ for the wavepacket states defined in Equation~\eqref{eq:initialstate} is shown in Figure~\ref{fig:initialwpckt} for different values of the parameters $p_0$ and $\delta$, the initial momentum and momentum spread respectively. In all cases, $\langle \phi (x) \rangle$ is peaked around $x=L/2$, indicating that this is the most probable separation between the two wavepackets. Despite preparing Gaussian wavepackets, the expectation value of $\langle \rho(x) \rangle$ does not take a purely Gaussian form, but instead exhibits oscillations within a Gaussian envelope. This pattern is caused by the interference between the amplitude of a wavepacket with momentum $-p_0$ located at a distance $L/2$ from its corresponding partner with momentum $p_0$, and the amplitude of a wavepacket with momentum $-p_0$ located at a distance $-L/2$ from its corresponding partner with momentum $p_0$. Such interference between wavepackets that are well separated in space is possible because $\langle \rho (x) \rangle $ is \emph{not} a local measurement. In the left panel of Figure~\ref{fig:initialwpckt}, we highlight the dependence of $\langle \rho (x) \rangle $ on $\delta$, and show explicitly how the envelope broadens as $\delta$ increases. In the right panel, we explore the dependence of $\langle \rho (x) \rangle $ on $p_0$, and observe that the spacing between peaks in the interference pattern scales as $\pi/p_0$. 


\subsubsection{Real-time evolution of scattering states}

\begin{figure}[t]
	\centering
		\includegraphics[width=\textwidth]{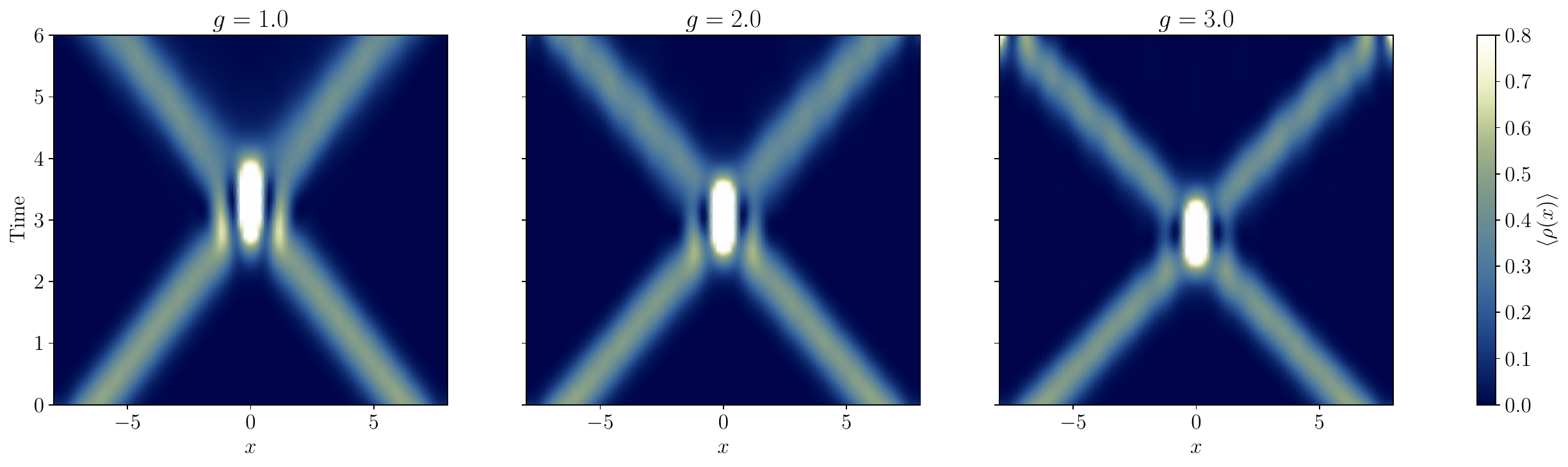}
		\caption{Movement and dispersion of wavepackets for varying coupling strengths using 10 qubits on a quantum emulator. For each coupling, the initial wavepackets are prepared with momentum $p_0 = 2.5$ and momentum spread $\delta = 0.75$. The interacting scattering states are generated via adiabatic state preparation with a ramp time of one time unit, and the Trotterised time evolution is performed with a time step $\delta t = 0.01$.}
	\label{fig:coupling}
\end{figure}

Now equipped with a means of extracting data from the scattering event we turn to the real-time evolution of the wavepackets in (1+1)-dimensional $\phi^4$ scalar field theory. Wavepackets with separation $L/2$ and momentum spread $\delta= 0.75$ are adiabatically prepared on a circle of length $L=16$, where we have set $M=1$. This is achieved by first constructing wavepackets at positions 0 and $L/2$ in the free theory, evolving under the free-field Hamiltonian for 1.5 time units to displace the wavepackets, and then adiabatically ramping up the coupling to prepare the scattering states in the interacting theory. A total ramp time of $\tau=1.0$ is used for this adiabatic evolution.

To study the wavepacket dynamics in the interacting theory, we simulate scattering events both varying the coupling strength and the initial momenta of the wavepackets. Each simulation is performed using Trotterised time-evolution, with a small time step of $\delta t = 0.01$, chosen to suppress Trotter error. We enforce a truncation of $n_q=10$. The scattering simulations have been performed on a quantum emulator, which simulates a fully fault-tolerant quantum device.

Figure~\ref{fig:coupling} shows the real-time dynamics of the two wavepackets scattering in the interacting field theory for varying coupling strengths $g=1.0$, 2.0 and 3.0. The wavepackets are prepared with an initial momentum $p_0 = 2.5$. As the coupling increases, we observe a shift in the interaction point to earlier times as the coupling increases. This is a consequence of the fact that, in $\phi^4$ scalar field theory, increasing the coupling effectively reduces the physical mass of the theory. As one approaches the critical point of the Ising phase transition, the theory moves towards a conformal field theory (CFT), where the mass gap vanishes. This reduction in mass leads to faster propagation, and thus an earlier interaction time at higher couplings~\cite{Rychkov:2014eea}.

\begin{figure}[t]
	\centering
		\includegraphics[width=\textwidth]{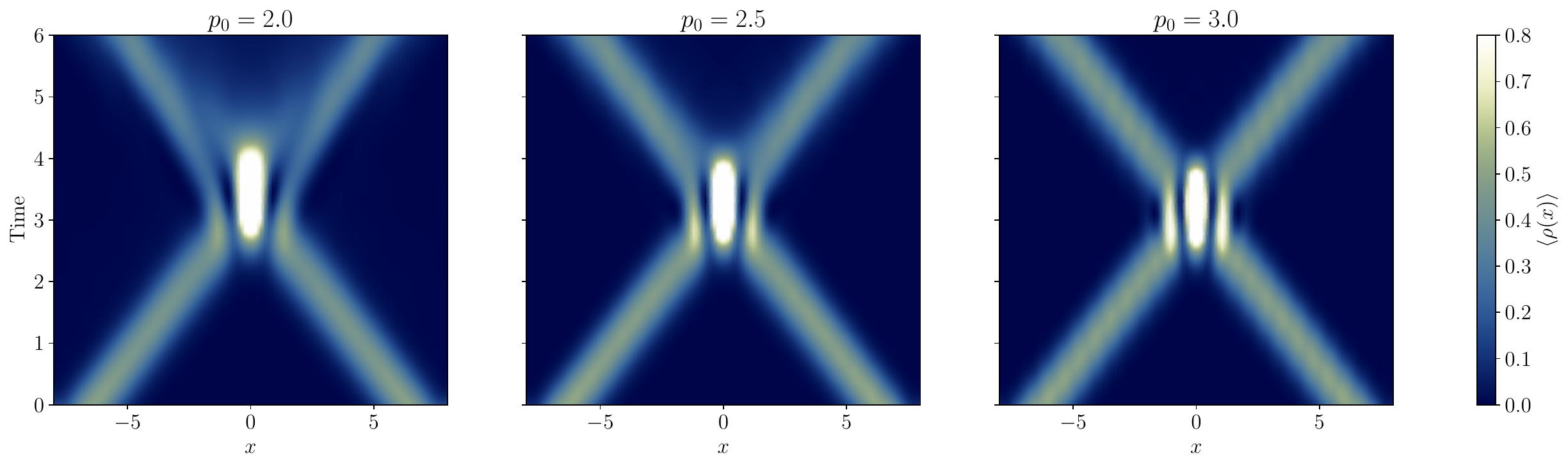}
		\caption{Movement and dispersion of wavepackets for varying initial momenta using 10 qubits on a quantum emulator. For each case, the wavepackets are prepared in the interacting picture with coupling $g=1.0$ via adiabatic state preparation with a ramp time of one time unit. Trotterised time evolution is performed with a time step $\delta t = 0.01$.}
	\label{fig:momenta}
\end{figure}

Another useful validation of the model is to vary the initial momenta of the wavepackets while keeping the coupling fixed. Figure~\ref{fig:momenta} shows the real-time dynamics of the two prepared wavepackets in the interacting theory for different initial momenta $p_0 = 2.0$, 2.5 and 3.0, at a fixed coupling strength of $g=1.0$. As the momentum increases, we observe that the spacing between fringes in the interference pattern decreases. This behaviour is consistent with the observation in Figure~\ref{fig:initialwpckt}, where the fringe spacing scales as $\pi/p_0$. In addition to the interference structure, the interaction vertex shifts to earlier times as $p_0$ increases. This is because larger initial momenta correspond to higher wavepacket velocities, resulting in faster propagation and an earlier point of interaction.

\begin{figure}[t]
	\begin{center}
		\includegraphics[width=\columnwidth]{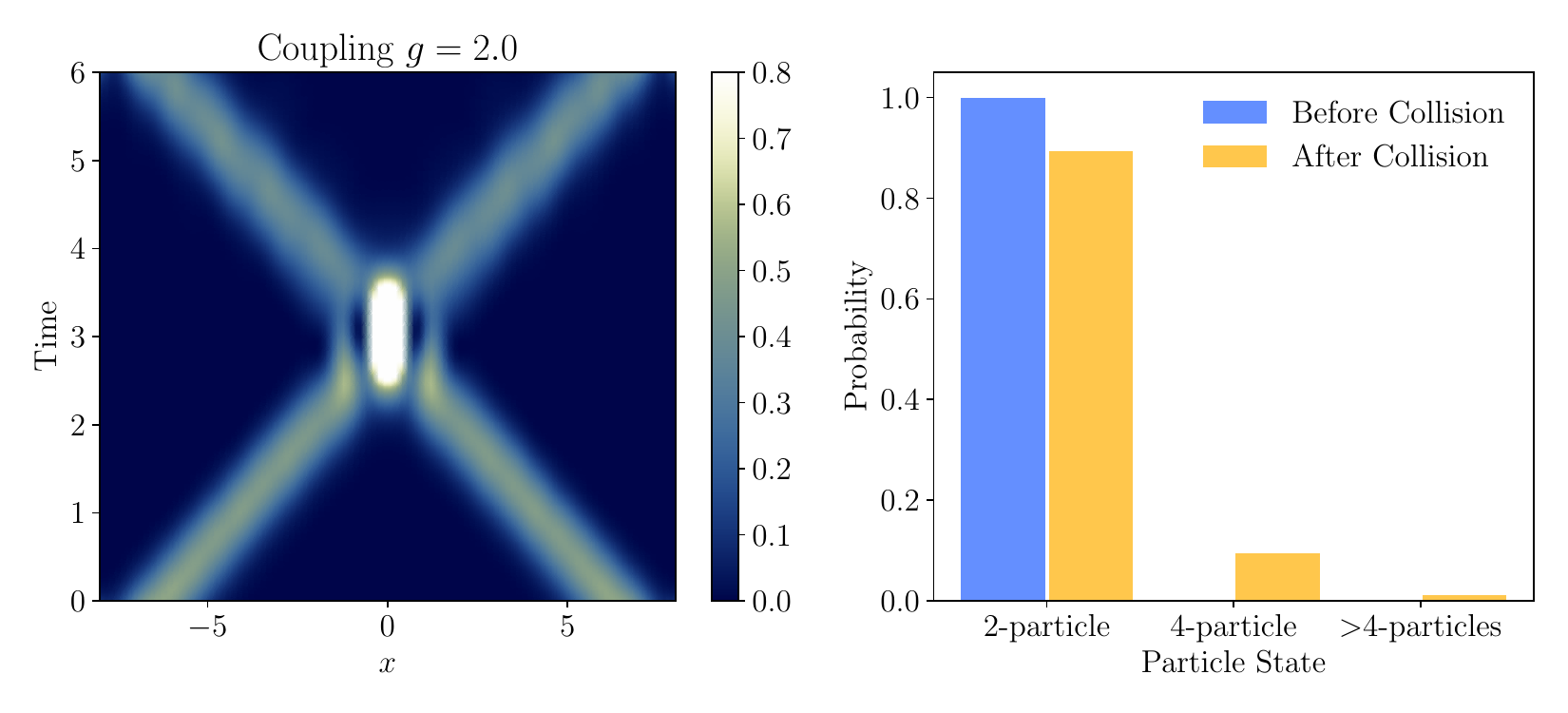}
	\end{center}
	\caption{Scattering-driven particle production. The two wavepackets are initialised with momentum $p_0=2.5$ and spread of $\delta=0.75$, evolved under a coupling strength of $g = 2.0$ with truncation of $n_q = 10$. The left panel shows the time evolution of the wavepacket density during the scattering process. The right panel shows the particle number distribution before $(t= 0.0)$ and after $(t=6.0)$ the collision, showing production of higher occupation number states after the collision. The simulation was performed with a Trotter step of $\delta t = 0.01$. }\label{fig:particleProd}
\end{figure}

An essential feature of scattering processes in interacting quantum field theories is the production of new particle states as a result of the interaction. To probe this phenomenon in the interacting $\phi^4$ theory, we consider a system of wavepackets with initial momenta $p_0 = 2.5$ and a momentum spread of $\delta = 0.75$, evolved with coupling strength $g=2.0$. As in previous simulations, we enforce a truncation of $n_q = 10$. Figure~\ref{fig:particleProd} shows the real-time dynamics of the scattering process alongside the corresponding occupation probabilities in the free-field basis before and after the collision. Prior to the interaction, as expected the system is composed entirely of two-particle states. However, after the collision, at time $t=6.0$, we observe a non-zero probability for higher particle-number occupation states, indicating that particle production has occurred due to the interaction. This is a clear indicator that quantum device can capture the non-trivial real-time dynamics of the QFT.

Together, Figures~\ref{fig:coupling}, \ref{fig:momenta} and \ref{fig:particleProd} illustrate the rich dynamics of real-time scattering events in (1+1)-dimensional $\phi^4$ scalar field theory accessible through quantum simulation. Furthermore, these non-trivial dynamics of the quantum field theory have been captured on only 10 qubits on the quantum emulator, highlighting the efficiency of the HT framework for quantum simulation, when combined with adiabatic state preparation and Trotterised time-evolution. These results not only serve as a validation of the simulation methodology but also demonstrate the potential of near-term quantum devices to explore scattering in complex quantum field theories. 


\section{Preparing scattering states on trapped ion quantum computers}\label{sec:QCRunSection}

To validate the Hamiltonian Truncation (HT) framework for quantum simulation of (1+1)-dimensional $\phi^4$ scalar field theory, we test the construction of interacting wavepackets using adiabatic state preparation~\cite{farhi2000quantumcomputationadiabaticevolution, RevModPhys.90.015002, 10.1063/1.2798382, 10.1063/1.4748968, Jordan:2011ci} on a noisy intermediate-scale quantum (NISQ) device. For this study, we use the IonQ Aria 1 quantum computer, a 25-qubit trapped-ion device based on $^{171}\textrm{Yb}^+$ ions~\cite{Wright_2019}. The trapped-ion platform is particularly well suited to the HT formalism due to its all-to-all qubit connectivity, which allows two-qubit gate operations to be applied across any two qubits. This is particularly useful for simulating the Hamiltonians that naturally arise in the HT approach, which lack manifest locality. In contrast, superconducting qubit devices~\cite{kjaergaard2020superconducting} typically have limited connectivity and require additional overhead to implement two-qubit operations. Furthermore, the IonQ Aria 1 device benefits from long coherence times, with $T_1$ and $T_2$ times of order $10-100$s and $1$s, respectively~\cite{IonQ1, IonQ2}, thus making them suitable for simulating scattering events which typically have large circuit depths. 

We construct scattering configurations in the interacting theory by first constructing wavepackets in the free theory and then adiabatically turning on the interaction, as outlined in Section~\ref{sec:adiabaticStatePrep}. As a proof-of-principle, we perform the state preparation on the quantum device, choosing an initial momentum of $p_0= 1.5$ with a momentum spread of $\delta = 0.75$, and prepare the wavepackets in the interacting theory with coupling strength $g= 2.0$, enforcing a truncation of $n_q = 4$. As before, we take the circle to have length $L=16$, and set the scalar mass to $M=1$. 

With this truncation, the free theory includes six two-particle states in the occupation number basis. By redefining the ordering of basis states, we map all two-particle states to be within the first eight basis elements. This relabelling enables the initial free-field state to be prepared on just three qubits instead of four, thereby reducing the circuit depth of the free theory wavepacket preparation from a circuit depth of 45 to 13, of which three gates are two-qubit \textsc{cnot} operations and six are virtual $R_Z$-rotations\footnote{On trapped-ion devices, it is not necessary to implement a physical $R_Z$ rotation by waiting for the qubit's phase to evolve in time. Instead, a virtual $R_Z$ operation can be performed by adjusting the phase of subsequent gate operations~\cite{IonQ2}.}. 

\begin{figure}
	\centering
	\includegraphics[width=\textwidth]{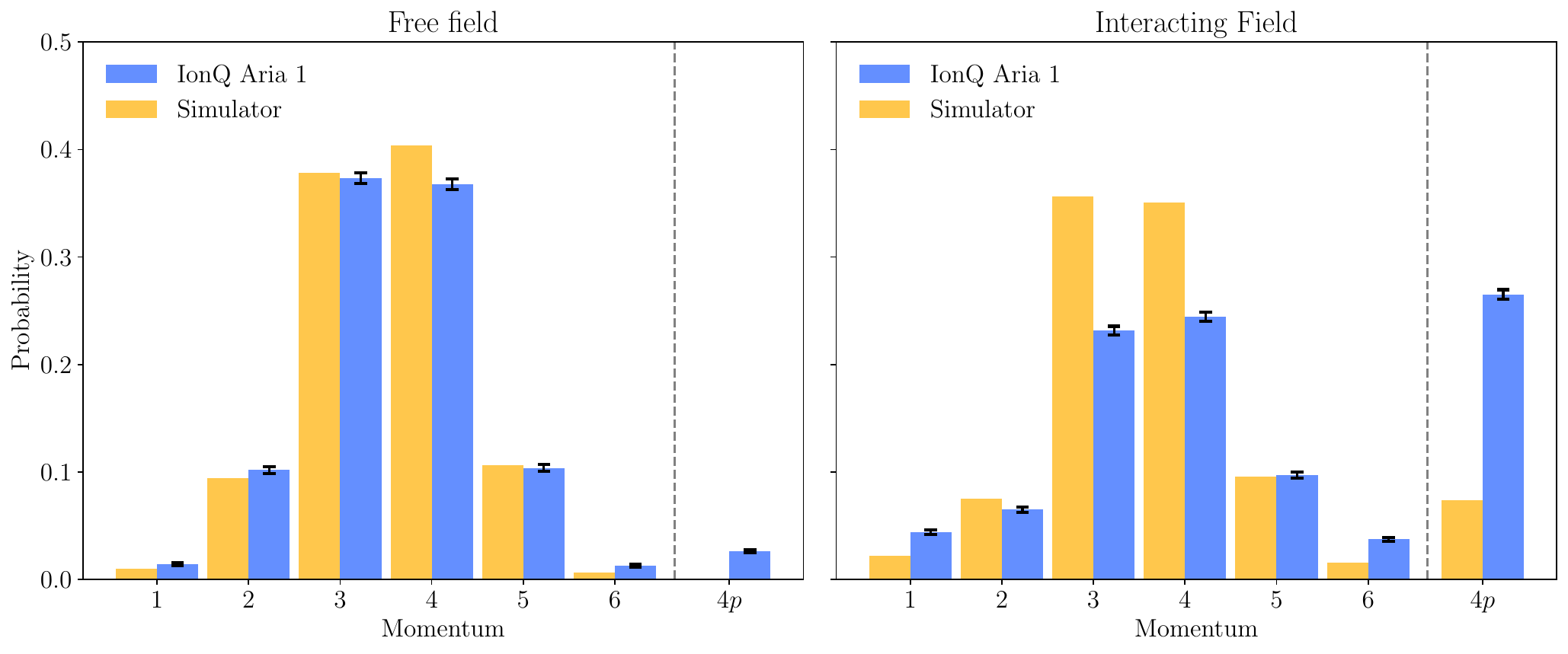}
	\caption{Occupation number probabilities for the preparation of free (left) and interacting (right) wavepackets, comparing the IonQ Aria 1 quantum computer against an ideal, noiseless quantum emulator. }\label{fig:statePrepFree}
\end{figure}

The left-hand panel of Figure~\ref{fig:statePrepFree} shows the occupation number probabilities for the free theory wavepackets obtained from the IonQ Aria 1 trapped-ion quantum computer, compared with results from a quantum emulator. The quantum device has been run for 10,000 shots, and statistical uncertainties are shown. The results demonstrate good agreement between the hardware and the emulator, although some leakage into four-particle states is observed. This discrepancy is likely due to gate errors on the device. The IonQ Aria 1 device has a reported one-qubit gate error rate of $0.05\%$ and a two-qubit gate error rate of $0.4\%$~\cite{IonQ2, PhysRevLett.125.150505}. In addition, the device has an associated state preparation and measurement (SPAM) error of $0.39\%$~\cite{Wright_2019, IonQ1, IonQ2}, which arises from inaccuracies in both initialising and measuring qubit states~\cite{PhysRevLett.113.220501}. Errors due to dephasing and energy relaxation, characterised by the $T_2$ and $T_1$ times respectively, are expected to be negligible in comparison. The cumulative effect of these noise sources contributes approximately a $2\%$ error in the simulation, accounting for the slight discrepancy between the distributions shown in Figure~\ref{fig:statePrepFree}.

The right-hand panel of Figure~\ref{fig:statePrepFree} shows the occupation number probabilities for the Gaussian wavepackets in the full, interacting theory, obtained from the IonQ Aria 1 trapped-ion quantum computer and compared with results from a quantum emulator. As before, the quantum data result from 10,000 shots, and statistical uncertainties are shown. The momentum distribution is prepared via adiabatic state preparation, in which the coupling is gradually turned on over a time $\tau$.  Here, we have chosen $\tau = 1.0$ to be long enough for the evolution to remain adiabatic, but short enough to prepare the wavepackets before the scattering takes place. Trotterised time evolution is performed with a chosen time step of $\delta \tau = 0.2$ to facilitate the simulation of the full adiabatic time-evolution on a NISQ device, whilst minimising the induced Trotter error. The circuit depth per Trotter step increases with the coupling strength; the total circuit depth of the adiabatic state preparation combined with the initial state preparation being 137, comprising 97 two-qubit gate operations. While the quantum results qualitatively follow the expected distribution, a significant deviation from the emulator is observed. This discrepancy is again attributed to gate and SPAM errors. The cumulative two-qubit gate error alone is $1 - (0.996)^{97} \approx 32\%$, accounting for the observed mismatch in Figure~\ref{fig:statePrepFree}.


\section{Resource Scaling and Comparisons with the Lattice}\label{sec:resourcesSec}

When assessing the feasibility of quantum simulations of scattering processes, two key resources are especially relevant: the number of qubits required and the total number of quantum gates needed to implement the simulation. To benchmark the demands of the Hamiltonian Truncation (HT) approach presented in the previous sections, we compare its resource requirements to those of the specific Hamiltonian lattice formulation used in the original works by Jordan, Lee, and Preskill (JLP) \cite{Jordan:2011ci,Jordan:2012xnu}. While alternative lattice formulations exist, this comparison focuses on the JLP variant to enable a direct and meaningful evaluation. In addition to quantifying resource scaling, we also highlight key qualitative differences between the two approaches, and put into context the trade-offs that are involved when choosing between them.

The JLP approach employs a Hamiltonian lattice formulation in which space is discretised into a one-dimensional chain of lattice sites, while time remains continuous. At each site, the scalar field is represented by an unbounded continuous variable, which must be discretised to make the Hilbert space finite-dimensional. This is accomplished by imposing a cutoff on the field amplitude and encoding the field using a discrete variable that can take one of $2^{n_q}$ evenly spaced values within the allowed range. The field at each lattice site is thus represented using a register of $n_q$ qubits. For a spatial interval of length $L$, and lattice spacing $a$, the total number of qubits required is then $N_q = n_q (L / a)$.

To compare the qubit requirements of the two approaches in a meaningful way, it is useful to consider the simulation of a scattering process with a fixed, high centre-of-mass energy $\sqrt{s}$. In the HT formulation, the total energy of the scattering state must lie below the ultraviolet cutoff used to truncate the Hilbert space. Similarly, in the lattice formulation, the maximum resolvable energy is set by the inverse of the lattice spacing, since momenta are bounded by the Brillouin zone. To enable a direct comparison, we therefore match the two schemes by requiring that the maximum scattering energy satisfies 
\begin{align} 
	\sqrt{s}\gtrsim E_\text{max} \approx 1/a\,.
\end{align} 
This identification allows us to express the total qubit count in the lattice formulation in terms of the physical energy scale being simulated, and thereby compare it directly to the corresponding requirements in the HT approach.

\begin{figure}[t]
	\centering
	\begin{subfigure}[t]{0.48\textwidth}
		\centering
		\includegraphics[width=\textwidth]{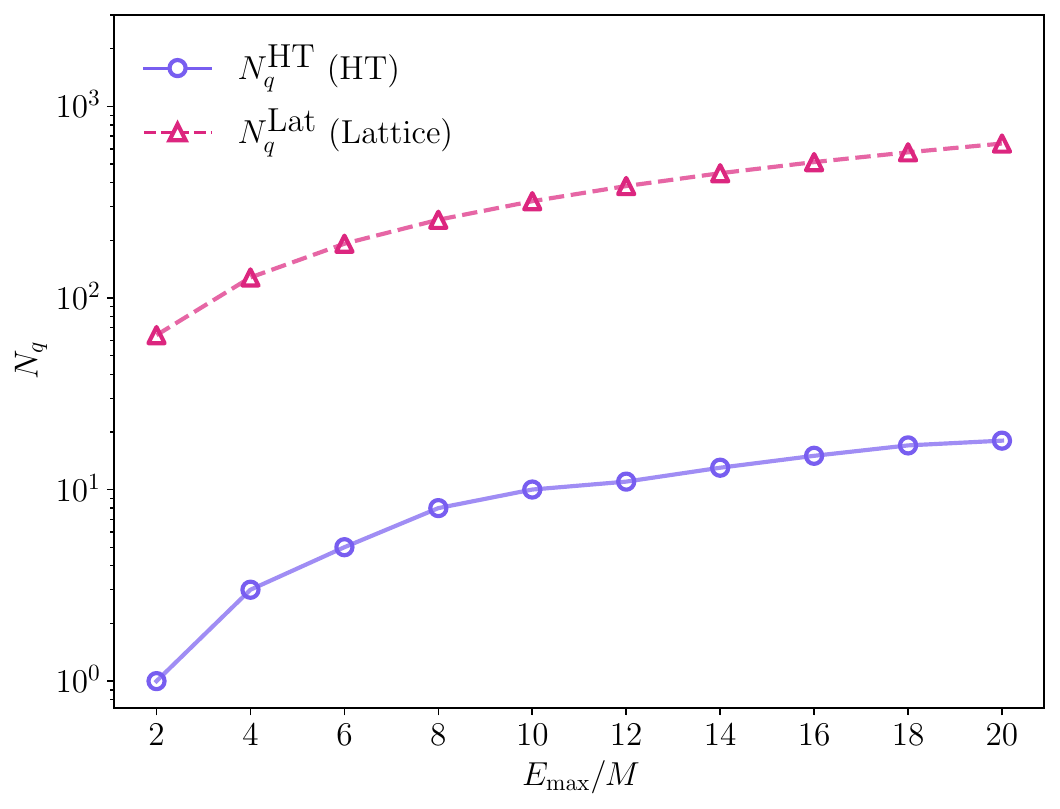}
	\end{subfigure}
	\begin{subfigure}[t]{0.48\textwidth}
		\centering
		\includegraphics[width=\textwidth]{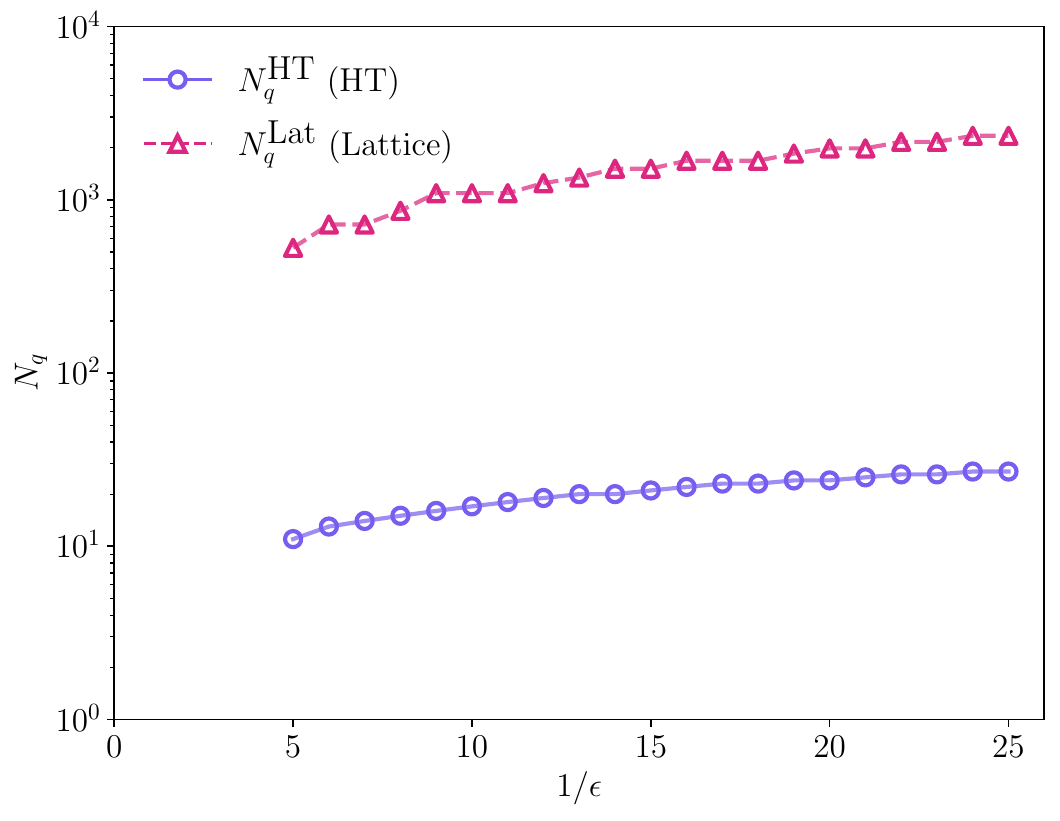}
	\end{subfigure}
	\caption{(Left) number of qubits required to simulate a scattering process in HT with maximum initial state energy $E_\text{max}$, compared with the number required to do the same using the Hamiltonian lattice approach, We have taken the number of qubits per lattice site to be $n_q=2$, and the volume to be $ML=16$. (Right) Number of qubits required to calculate the $2\rightarrow4$ scattering cross section with precision tolerance $\epsilon$ in both HT and lattice approaches. We have taken $g=M^2$ and $\sqrt{s}=5m$.}\label{fig:comparison}
\end{figure}

The left panel of Figure~\ref{fig:comparison} shows the total number of qubits required to simulate scattering processes up to a given maximum energy $E_\text{max}$, for both the Hamiltonian Truncation and lattice approaches. For reference, we have fixed the physical volume to $ML = 16$ in both cases, and taken $n_q = 2$ for the lattice calculation, corresponding to a local Hilbert space of 4 field values per site. The total number of qubits required in the lattice approach is then estimated using $N_q = n_q (ML) (E_\text{max}/M)$. The corresponding estimate for HT is obtained by directly computing the number of states in the truncated basis with $H_0$ eigenvalue less than $E_\text{max}$ for the given lattice volume. In both schemes, the total qubit count grows approximately linearly with $E_\text{max}$. However, the lattice approach exhibits a significantly steeper cost: across the plotted range, it requires nearly 40 times more qubits than the corresponding lattice calculation to access the same maximum energy.

In addition to estimating the qubit requirements for simulating processes at high energy, we also compare the resources needed to simulate a process with moderate, fixed energy but high precision. To this end, we consider $2 \rightarrow 4$ particle scattering at a centre-of-mass energy $\sqrt{s} = 5m$. This same process is also examined in Appendix~F of Reference~\cite{Jordan:2012xnu}, providing a useful point of comparison, and our estimate of the resources needed in the lattice approach will closely follow that work.  In the following discussion, we focus exclusively on uncertainties arising from the field theory truncations employed in each approach; we do not consider errors due to quantum hardware or gate-level noise.

In the lattice approach, the error introduced by discretising space can be estimated as $\epsilon\sim (pa)^2$ by looking, for instance, at the difference between continuum and lattice dispersion relations, implying a discretisation error of $\epsilon\sim 4(ma)^2$ for the $2\rightarrow4$ process. Analogously in HT, the error arising from the Hilbert space truncation can be analysed using an effective Hamiltonian formalism. This effective Hamiltonian acts within the truncated Hilbert space and is constructed to reproduce the low-energy spectrum of the full theory. See, for example, Reference~\cite{Cohen:2021erm} for a detailed construction of an effective Hamiltonian in the case of $\phi^4$ theory. Comparing the contribution to energies from the leading $O(g^2)$ correction in the effective Hamiltonian to the contribution from $H_0$, we estimate this truncation error to be
\begin{align}
	\epsilon \approx \frac{(4!)^2g^2}{4\pi E^2_\text{max} m^2}\,,
	\label{eq:hterr}
\end{align}
where factors of $4!$ arise from the normalisation for the quartic coupling, and the factor of $4\pi$ is motivated from NDA counting rules in $1+1$ dimensions \cite{Gavela:2016bzc}, and because the $O(g^2)$ correction to the effective Hamiltonian receives contributions from one--loop diagrams.

The finite volumes used for simulations of scattering constitute another relevant source of error. In \cite{Jordan:2012xnu}, the minimum volume needed to reduce this error to the same level as the discretisation error was estimated by requiring the forces between final state particles to be small so that
\begin{align}
	\epsilon \ge \frac{\partial V / \partial r |_{r=r_0}}{\partial V / \partial r |_{r=1/m}}\,,
	\label{eq:force}
\end{align}
where $r_0$ is the separation between any pair of $\phi$ particles in the final state of the scattering process. To ensure all four particles are adequately separated, even if they are not evenly spaced, we take the total volume to be $L=6r_0$. Here, $V(r)$ is the interparticle potential, which for large $r$ but lowest order in the coupling takes the form 
\begin{align}
	V(r) = -\frac{18g^2}{m^3}\frac{1}{\sqrt{\pi mr}}e^{-2mr}\,.
	\label{eq:interpotential}
\end{align}
In both the HT and lattice approaches, we use Equations~\eqref{eq:force} and \eqref{eq:interpotential} to determine numerically the minimum volume required to simulate $2\rightarrow 4$ scattering with error below $\epsilon$.

In the lattice approach, an additional source of error arises from the discretisation of the field variable. To control this, the number of qubits required per lattice site can be estimated as
\begin{align}
	n_q \approx \Bigg\lceil \log_2 \left(1+\frac{4\sqrt{s}}{m\pi}\left(1+\sqrt{\frac{L}{a\epsilon}}\right)^2\right)\Bigg\rceil\,.
	\label{eq:qubspersite}
\end{align}
This estimate may be somewhat conservative; it is often possible to reduce $n_q$ without compromising precision by tuning the discretisation parameters for the field variable at each site, as discussed in Reference~\cite{Klco:2018zqz}. Nonetheless, for the purposes of comparison with Reference~\cite{Jordan:2012xnu}, we adopt the expression in Equation~\eqref{eq:qubspersite}, taking $\sqrt{s}=5m$.

By combining the estimates in Equations~\eqref{eq:hterr} to \eqref{eq:qubspersite}, we compare the qubit requirements of the Hamiltonian Truncation and lattice approaches for varying precision tolerances $\epsilon$, as shown in the right panel of Figure~\ref{fig:comparison}. Our estimates confirm that significantly fewer qubits are required for a calculation of the $2\rightarrow 4$ particle scattering cross section for a given level of precision using HT than would be required for the lattice Hamiltonian approach.

\begin{figure}[t]
	\centering
	\includegraphics[width=0.5\textwidth]{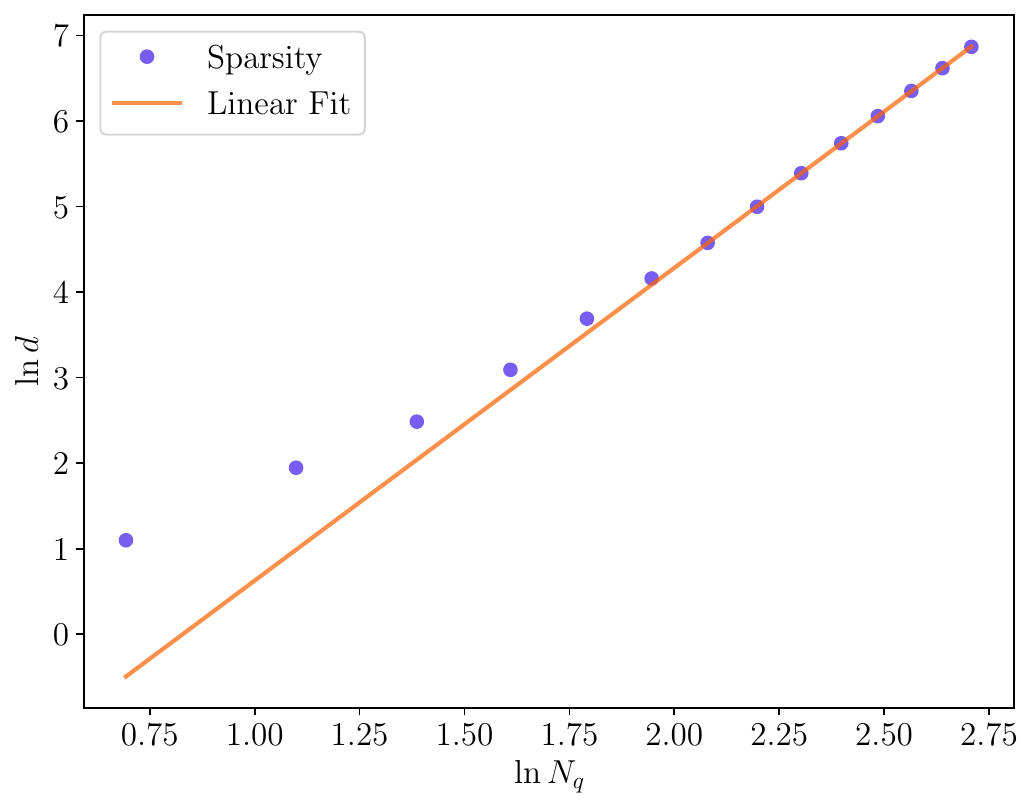}
	\caption{Log-Log plot showing the growth in the number of nonzero matrix elements in the truncated HT Hamiltonian with qubit number $N_q$. The quantity $d$ refers to the matrix sparsity, defined as the maximum, over all rows, of the number of nonzero elements in a given row. We find that $d\ll 2^{N_q}$ throughout the range, confirming that the HT Hamiltonian is sparse. In addition, the linear growth indicates that $d$ grows polynomially with $N_q$. The result of a linear fit to all but the lowest five points is the functional form $\ln d = 3.6651(63) \ln N_q - 3.028(15)$.}
	\label{fig:sparsity}
\end{figure}

The circuit depth and gate count required to simulate time evolution differ significantly between lattice and Hamiltonian Truncation (HT) approaches. This difference arises primarily due to differences in how locality is treated. On the lattice side, the Hamiltonian is explicitly local: each term typically involves fields or their conjugate momenta at a small number of neighbouring lattice sites. As a result, time evolution via the Suzuki–Trotter formula can be implemented with a number of quantum gates that grows polynomially with the number of lattice sites and with the number of qubits per site. This favourable scaling is a direct consequence of the locality of interactions, which limits the connectivity required in the quantum circuit. Foundational work on Hamiltonian simulation~\cite{Lloyd:1996aai,Jordan:2011ci,Jordan:2012xnu} has shown how this locality enables efficient digital quantum simulation schemes. 

By contrast, in the HT approach, the truncated Hilbert space is not built as a tensor product of local site-based spaces. Consequently, the locality of the original field theory is not manifest, and the Hamiltonian tends to contain couplings between widely separated basis states. This leads to a Pauli string decomposition of the time evolution operator, such as that appearing in Equation~(\ref{eq:utrott}), whose number of terms, and hence the number of required quantum gates scales exponentially with the number of qubits $N_q$. This exponential scaling applies to each Trotter step, making circuit depth a potential bottleneck in HT-based simulations when using Suzuki–Trotter methods.

Despite the exponential gate scaling with qubit number $N_q$, HT remains competitive because it can achieve a given target precision or energy cutoff with substantially fewer qubits than lattice-based methods. As a result, the total number of gates may remain lower in HT, even though its per-qubit scaling is less favourable. This advantage is most relevant for moderate or smaller problem sizes, where the benefits coming from the reduced qubit overhead are greatest relative to the costs from circuit depth scaling.

Furthermore, while HT Hamiltonians may lack manifest spatial locality, they nevertheless are often sparse. This opens the door to alternative quantum simulation algorithms that go beyond Suzuki–Trotter decompositions. Techniques based on quantum walks and block-encoded oracles, such as those developed in Refs~\cite{Berry:2013tiy,Berry:2015hst,Low:2016sck,Low:2016znh} offer better asymptotic scaling of gate depth with system size for sparse Hamiltonians. In the case of the HT formulation of $\phi^4$ theory used in this work, we confirm in Figure~\ref{fig:sparsity} that the truncated Hamiltonian is indeed sparse, with the number of nonzero entries growing only polynomially with the number of qubits, whilst the total number of states in the truncated basis grows exponentially. This raises the possibility of implementing time evolution with quantum circuits whose depth scales polynomially rather than exponentially with the problem size. Recent studies have demonstrated the applicability of these techniques to sparse Hamiltonians obtained using related HT methods~\cite{Liu:2020eoa,Kirby:2021ajp,Kreshchuk_2022}.

However, in general post-Trotter algorithms require significant gate overheads and additional ancilla qubits, which can make them impractical for near-term quantum hardware. Their advantages only become manifest at large problem sizes, beyond the current capabilities of quantum devices; see for example the comparisons in \cite{Hariprakash:2023tla,Hardy:2024ric, hanada2025exponentialimprovementquantumsimulations}. In future work, it would be interesting to develop explicit applications of these more advanced time-evolution algorithms within the HT framework, particularly as quantum hardware continues to improve and allows for larger-scale simulations.

Although $\phi^4$ theory in $1+1$ dimensions represents a particularly simple example, it nevertheless illustrates many of the resource advantages offered by Hamiltonian Truncation (HT) relative to standard Hamiltonian lattice formulations. In our implementation, we construct the truncated Hamiltonian directly within the subsector of the full Hilbert space that is parity even and has total momentum zero. This targeted truncation leads to a dramatic reduction in the number of qubits required for simulation. As shown in Figure~\ref{fig:comparison}, this economy of representation contrasts sharply with lattice-based approaches, where such global symmetry constraints are not automatically incorporated into the construction of the Hamiltonian. Instead, lattice simulations typically encode the full local Hilbert space at each site, leading to significantly larger qubit demands even for modest system sizes. While it is possible in principle to impose global symmetry constraints on the lattice (see for instance \cite{PhysRevA.98.032331}), doing so generally breaks the structure of the Hilbert space as a tensor product of local Hilbert spaces for each site. As a result, the locality of interactions, a central strength of the lattice formulation, is lost, complicating both the construction of and time evolution using the resulting Hamiltonian.

More generally, HT offers a number of additional structural advantages for quantum simulation. It enables the formulation of theories that preserve exact continuum symmetries, such as translations and rotations, and, where applicable, more intricate structures like chiral or supersymmetry\footnote{Examples of HT studies on spacetimes with greater than 2 dimensions, in which exact rotational invariance is preserved by the truncation include Refs.~\cite{Hogervorst:2014rta,Hogervorst:2018otc}. Supersymmetric theories have been studied using HT with lightcone quantisation in Refs~\cite{Matsumura:1995kw,Fitzpatrick:2019cif}.}. In addition, HT can be applied to QFTs that lack a known or tractable lattice regularisation, including certain strongly coupled conformal field theories perturbed by relevant operators.

These advantages come with trade-offs that stem from the non-local nature of the HT truncation. States are selected based on total energy (more precisely, their $H_0$ eigenvalue), with no consideration of how that energy is distributed in space. As a result, non-local counter-terms can be required to ensure proper renormalisation. Although the resulting Hamiltonians may remain sparse in the truncated basis, their non-local structure leads to unfavourable circuit depth scaling for time evolution protocols based on naive Suzuki–Trotter decompositions. The choice between HT and lattice methods therefore reflects a fundamental trade-off between symmetry and locality: HT offers a flexible framework for incorporating symmetries, efficiently reducing the Hilbert space dimensionality, while the lattice approach enforces locality from the outset, facilitating scalable simulation strategies.


\section{Summary and conclusions}\label{sec:conclusion}

We present a quantum computing framework for simulating real-time scattering processes in $(1+1)$-dimensional $\phi^4$ scalar field theory using Hamiltonian Truncation (HT). Employing this framework, we efficiently approximate the Hilbert space of the quantum field theory (QFT), thereby significantly reducing the number of qubits required compared to traditional lattice-based approaches. Moreover, the ground state of the free QFT is constructed such that it corresponds directly to the ground state of a qubit-based device. This identification obviates the need for complex ground state preparation routines and reduces the overall circuit depth of the algorithm. Consequently, the preparation of scattering states within the HT framework is simplified: wavepackets can be constructed efficiently in the free theory and adiabatically evolved to the interacting theory.

We validate the method on both quantum emulators and noisy intermediate-scale quantum (NISQ) hardware, capturing physical aspects of the theory, such as wavepacket dynamics, interference patterns, and particle production post-scattering. These simulations highlight the potential of HT for capturing complex, non-perturbative phenomena in quantum field theories.

To assess the practical applicability of the framework, we implement our state preparation protocol on the IonQ Aria 1 quantum computer, a NISQ-era trapped-ion device. We demonstrate how a reordering of the basis states enables a reduction in circuit depth, permitting high-fidelity preparation of free-theory wavepackets. Scattering states in the interacting theory are then prepared via adiabatic state preparation. Although the results for the interacting theory show qualitative agreement with simulations on a fault-tolerant quantum emulator, significant discrepancies remain. In particular, the experiments exhibit noticeable leakage into unintended multi-particle occupation states, attributed primarily to gate errors on the NISQ device. Thus, while we have demonstrated that the core principles of the HT method can be executed on a real quantum device, substantial improvements in quantum hardware fidelity will be required to realise its full potential.

Our analysis also includes a detailed resource comparison between the HT and lattice-based methods, demonstrating that the HT approach substantially reduces qubit requirements. Specifically, for fixed physical parameters such as centre-of-mass energy or target precision, the HT method typically requires an order of magnitude fewer qubits than lattice Hamiltonian approaches, often by a factor of 40 within the energy ranges considered. This economy arises because HT targets only the physically relevant sector of the Hilbert space, whereas lattice simulations must encode the entire local field space at each lattice site. However, we find that circuit depth grows exponentially with problem size, due to the lack of manifest locality in truncated Hamiltonians. This scaling poses a significant challenge, particularly in light of the constraints of current quantum hardware. Exploring alternative quantum simulation methods, such as quantum walks or block-encoding algorithms, could provide promising avenues for future research to improve depth scalability.

While the HT approach offers promising theoretical and computational advantages, its practical application on current-generation NISQ hardware remains challenging. Nevertheless, our results suggest clear pathways forward: reducing noise and improving coherence in quantum hardware, optimising state preparation and time evolution algorithms, and further developing advanced quantum simulation techniques. These steps will be crucial for fully exploiting HT for quantum computing to address computationally challenging quantum field theory simulations.
	
\vspace{0.5cm}
\noindent{\textit{\textbf{Acknowledgments} This project was funded and supported by the UK National Quantum Computer Centre [NQCC200921], which is a UKRI Centre and part of the UK National Quantum Technologies Programme (NQTP). We would like to thank the NQCC and specifically Laura Martin for their technical support. We acknowledge the use of IonQ quantum devices through the AWS Braket cloud server. MW has been partially supported by STFC consolidated grant ST/X000664/1 and an IPPP Associateship.}

\bibliographystyle{inspire}
\bibliography{refs}{}

\end{document}